\documentclass[12pt]{iopart}

\usepackage{iopams}
\usepackage{setspace}

\usepackage{amsfonts}   
\usepackage{amssymb}
\usepackage{graphicx}   

\usepackage{url}

\newcommand{\diag}{diagonalization}  
\newcommand{\bh}{Bose-Hubbard}
\newcommand{\lan}{Lanczos}
\newcommand{\ultra}{ultracold}
\begin{document}

\title{Exact diagonalization: the Bose-Hubbard model as an example}
\author{J M Zhang and R X Dong}
\address{Institute of Physics, Chinese Academy of Sciences, Beijing
100080, China}

\ead{jmzhang@aphy.iphy.ac.cn}

\begin{abstract}
We take the \bh\ model to illustrate exact \diag\ techniques in a
pedagogical way. We follow the road of first generating all the
basis vectors, then setting up the Hamiltonian matrix with respect
to this basis, and finally using the \lan\ algorithm to solve low
lying eigenstates and eigenvalues. Emphasis is placed on how to
enumerate all the basis vectors and how to use the hashing trick to
set up the Hamiltonian matrix or matrices corresponding to other
quantities. Although our route is not necessarily the most efficient
one in practice, the techniques and ideas introduced are quite
general and may find use in many other problems.
\end{abstract}


\submitto{\EJP}


\section{Introduction}
Among the various analytical and numerical approaches to strongly
correlated systems, numerical exact \diag\ takes a unique position.
It is not burdened by any assumptions or approximations and thus
provides unbiased benchmarks for other analytical and/or numerical
approaches \cite{lin_90}. It is also appealing in its conceptually
simple and straightforward nature. The basic idea, to set up the
Hamiltonian matrix in some basis and thus reduce a physical problem
to a purely mathematical one, is readily accessible to a senior
undergraduate student.

However, possibly due to some technical subtleties, exact \diag\ is
not accounted for in detail in existing textbooks on computational
physics. It is the aim of this paper to illustrate these tricks and
promote teaching and using of exact \diag. To make the discussion
concrete, we take the \bh\ model as an example. This model is chosen
because of its relevance to the currently active field of \ultra\
atom physics \cite{blochr}. It has been realized with \ultra\ atoms
in an optical lattice and the celebrated Superfluidity-Mott
insulator (SF-MI) transition has been observed experimentally
\cite{jacksch,bloch}. We will use exact \diag\ to get a glimpse of
this quantum phase transition.

One common misconception, according to the experience of the
authors, is that in doing numerical exact \diag, one solves all the
eigenvalues and eigenvectors of the Hamiltonian by some algorithm.
This ideal case is actually neither possible nor necessary in many
cases as long as the dimension of the Hilbert space $D$ gets large.
It is impossible since to reduce a Hermitian matrix $H$ in the form
$U\Lambda U^\dagger$, with $\Lambda$ being a diagonal real matrix
and $U$ a unitary matrix, it would take time on the order of
$O(D^3)$ and memory space on the order of $O(D^2)$. With a moderate
value $D=100\,000$, the memory needed is over 10 GB, far beyond that
of a typical desktop computer, needless to say the time cost. It is
also unnecessary since physically, in many cases, the most relevant
eigenstates are the ground state and low lying excited states. High
excited states, due to the Boltzmann factor, contribute little to
the thermodynamics of the system in low temperatures.

In view of the considerations mentioned above, one can fully
appreciate the value of the \lan\ algorithm \cite{lanczos}. This
algorithm belongs to the iterative category for solving eigenvalue
problems. As the iteration goes on, the estimated eigenvalues and
eigenvectors converge quickly. Especially, the extremal eigenvalues
and eigenvectors converge first. Usually, with an iteration time
$m\ll D$, the ground state and several low excited states converge
to machine precision. In a certain sense, the \lan\ algorithm is a
tailor-made algorithm for solving the ground state and/or low lying
excited states of a Hamiltonian. It provides exactly what we need
for us, no more no less.

As far as we know, all exact \diag s are based on the Lanczos
algorithm and its variants. Since this algorithm has become a
standard topic in textbooks on numerical matrix theory
\cite{trefethen} and since there are many monographes \cite{mono}
devoted to this algorithm and also several very readable
introductions \cite{lin_93,springer}, here in this article, we would
not go into the details of this algorithm. We will just invoke some
packages based on Lanczos algorithm and use it as the final stroke.

On the contrary, our emphasis is placed on some other techniques
which we believe are involved in all kinds of exact \diag s in a
wide variety of contexts. The road map we will take is perhaps the
most natural one---first enumerate all the basis vectors, then set
up the Hamiltonian matrix in this basis, and finally invoke the
Lanczos algorithm to solve the desired eigenstates. In each step, we
will explain the tricks in detail. We would like to mention that in
practice, usually the Hamiltonian matrix is not explicitly set up
beforehand, instead the action of the Hamiltonian on a wave vector
is done ``on the fly'' \cite{lin_93,springer}. This is consistent
with the philosophy of iterative method---the matrix-vector
multiplication is all what we need and its internal workings are of
no concern. Though this is of use for saving the memory, we would
not introduce it here for our pedagogical purposes.

\section{The model and its symmetries}
We begin by describing the one dimensional Bose-Hubbard model and
its symmetries. The Hamiltonian is
\begin{equation}\label{bhm}
    \hat{H}=-J\sum_{\langle i j \rangle} (a_i^\dagger a_j
    +a_j^\dagger a_i)+\frac{U}{2}\sum_{i=1}^M \hat{n}_i(\hat{n}_i -1)£¬
\end{equation}
where $a_i^\dagger$ ($a_i$) creates (annihilates) a particle on the
site $i$ and $\hat{n}_i=a_i^\dagger a_i$ counts the particle number
on that site. The first term proportional to $J$ is the kinetic part
of the Hamiltonian ($\hat{H}_{kin}$) and describes particle hopping
between adjacent sites (in the sum, $\langle ij \rangle\equiv
\langle ji \rangle$). The second term is the interaction part
($\hat{H}_{int}$) and is due to the particle-particle interaction,
the strength of which is characterized by the parameter $U$. The
Bose-Hubbard model has been realized with ultracold boson atoms in
an optical lattice \cite{bloch}. Moreover, in this system, the
parameters $J$ and $U$ can be conveniently adjusted by various
means, e.~g. the Feshbach resonance or just changing the intensity
of the laser beams.

The Hamiltonian $\hat{H}$ possesses several symmetries. The first
one is the $U(1)$ symmetry, which is associated with the
conservation of the total atom number $\hat{N}=\sum_{i=1}^M
\hat{n}_i$. The Hamiltonian is invariant under the transform
$(a_i^\dagger,a_i)\rightarrow (a_i^\dagger e^{i\theta},
a_ie^{-i\theta})=e^{i\hat{N}\theta}(a_i^\dagger,a_i)
e^{-i\hat{N}\theta}$ for $\forall$ $\theta \in \mathbb{R}$. The
second one is the translation symmetry. The Hamiltonian is invariant
under the transform $(a_i^\dagger,a_i)\rightarrow
(a_{i+1}^\dagger,a_{i+1})$, if the periodic boundary condition is
imposed. This symmetry is associated with the conservation of the
total quasi-momentum of the system
\begin{equation}\label{momentum}
\hat{K}\equiv \sum_{q=0}^{M-1} \left(\frac{2\pi q}{M}\right)
b_q^\dagger b_q \pmod {2\pi},
\end{equation}
where the operator
\begin{equation}\label{bq}
    b_q^\dagger=\frac{1}{\sqrt{M}}\sum_{j=1}^M
e^{i(j\cdot 2\pi q/M)}a_j^\dagger
\end{equation}
creates a particle in the Bloch state with quasi-momentum $2\pi
q/M$. Actually, the transform above is done as
$e^{-i\hat{K}}(a_i^\dagger,a_i)e^{i\hat{K}}=(a_{i+1}^\dagger,a_{i+1})
$. In terms of $(b_q^\dagger,b_q)$, the Hamiltonian $\hat{H}$ is
rewritten as
\begin{eqnarray}\label{h momentum}
\fl  \quad\quad\quad \hat{H} =-2J\sum_{q=0}^{M-1} \cos(2\pi
q/M)b_q^\dagger b_q
    + \frac{U}{2M}\sum_{q_1,q_2=0}^{M-1}\sum_{q_3,q_4=0}^{M-1} b_{q_1}^\dagger b_{q_2}^\dagger b_{q_3}
    b_{q_4}\delta_{q_1+q_2,q_3+q_4},
\end{eqnarray}
where the Dirac function is defined as
\begin{equation}
\label{cases}
\delta_{q_1+q_2,q_3+q_4}=\cases{1 & if $q_1+q_2 \equiv q_3+q_4 \pmod {M} $ \\
0 & otherwise  \\}
\end{equation}
It is then clear that $\hat{K}$ is conserved. The third symmetry is
the reflection symmetry. The Hamiltonian is also invariant under the
transform $(a_i^\dagger, a_i)\rightarrow (a_{M-i}^\dagger, a_{M-i})$
\cite{identify}, or in terms of $(b_q^\dagger,b_q)$,
$(b_q^\dagger,b_q)\rightarrow (b_{-q}^\dagger,b_{-q})$. Combination
of the translation and reflection symmetries indicates that the
Bose-Hubbard model has the $D_M$ symmetry, the symmetry of a
equilateral polygon with $M$ vertices. This is plausible if we
envisage that the $M$ sites are placed equidistantly on a circle.

Therefore, the Bose-Hubbard model is of $U(1)\otimes D_M$ symmetry.
It is desirable to decompose the total Hilbert space into subspaces
according to the irreducible representations of this group. The
Hamiltonian cannot couple two states belonging to two different
irreducible representations and thus is block (partially)
diagonalized \cite{springer,jafari}. Analytically, it can be proven
that the ground state $|G \rangle$ of $\hat{H}$ belongs to the
identity representation of $D_M$ \cite{ground state}. Therefore, as
far as the ground state is concerned, we only need to seek it in a
subspace where all the basis vectors are of a definite atom number
[thus belong to a definite representation of $U(1)$] and are
invariant under all rotations and reflections.

We first restrict to the space $\mathcal {H}$ with total atom number
being $N$. The dimension of this space is found to be
\begin{equation}\label{dimension}
    D=\frac{(N+M-1)!}{N!(M-1)!},
\end{equation}
which grows explosively with the system size. For fixed filling
factor $N/M=1$, $D=24\,310$ for $M=9$, and it grows to $D=352\,716$
for $M=11$, and further to $D=5\,200\,300$ for $M=13$. We may divide
this space into $M$ smaller subspaces according to the eigenvalues
of $\hat{K}$. The ground state, being translationally invariant,
falls in the subspace $\mathcal {H}_0$ with $K=0$, whose dimension
$D_0$ is approximately $D/M$. Actually, $D_0=2\,704$, $32\,066$, and
$400\,024$ in the case of $N=M=9$, $11$, and $13$, respectively (see
reference~\cite{k0}). The subspace $\mathcal {H}_0$ can be further
divided into two subspaces according to the two representations of
the reflection group $\{I,\sigma \}$. The ground state, being
invariant under reflection, belongs to the subspace $\mathcal
{H}_0^+$, where the superscript means all the basis vectors yield a
plus sign under reflection. The dimension $D^+_0$ of $\mathcal
{H}_0^+$ is nearly half of $D_0$. Actually, $D_0^+=1\,387$,
$16\,159$, and $200\,474$ respectively in the three cases above. The
reduction of the dimension from $D$ to $D_0$ and again to $D_0^+$
promises a reduction of computation, especially, a reduction of
memory needed.

Indeed, when memory is limited, it is necessary to work within the
subspace $\mathcal {H}_0$ (or even $\mathcal {H}_0^+$) and with the
Hamiltonian in equation~(\ref{h momentum}) (so we do in the $N=M=13$
case below). However, to simplify the discussion and focus on
essential techniques, we will still work within the space $\mathcal
{H}$ and with the Hamiltonian in equation~(\ref{bhm}). In fact,
working in the subspaces $\mathcal {H}_0$ or $\mathcal {H}_0^+$
requires a bit more effort in coding and will be left as exercises.

\section{Basis vectors generation}\label{basis gen}
A natural basis is the occupation number basis
$\{|n_1,n_2,\ldots,n_M\rangle \}$ which are defined as
\begin{equation}\label{ocuupation}
\hat{n}_i |n_1,n_2,\ldots,n_M\rangle= n_i |n_1,n_2,\ldots,n_M\rangle
\end{equation}
with $n_i \geq 0$. In the subspace with a fixed total particle
number $N$, we have the constraint: $\sum_{i=1}^M n_i=N$. We need to
enumerate all the basis vectors satisfying this constraint. One
naive idea is to write down a piece of code with a $M$-fold loop:
\begin{verbatim}
for n_1=0:N
  for n_2=0:N-n_1
    for n_3=0:N-n_1-n_2
       ...
    end
  end
end
\end{verbatim}
This approach, though workable, has two apparent drawbacks. First,
the number of loops depends on the number of sites and hence the
code is inflexible. Second, when coding with tools such as MATLAB
which is inefficient in dealing with loops, the efficiency would be
low.

Here we prescribe one way to bypass these difficulties. To this end,
we first note that it is possible to rank all the basis vectors
$|n_1,n_2,\ldots,n_M \rangle$ in lexicographic order \cite{order}.
For two different basis vectors $|n_1,n_2,\ldots,n_M \rangle$ and
$|\bar{n}_1,\bar{n}_2,\ldots,\bar{n}_M \rangle$, there must exist a
certain index $1 \leq k \leq M-1$ such that $n_i=\bar{n}_i$ for $1
\leq i \leq k-1$ while $n_k \neq \bar{n}_k$. We say
$|n_1,n_2,\ldots,n_M \rangle$ is superior (inferior) to
$|\bar{n}_1,\bar{n}_2,\ldots,\bar{n}_M \rangle$ if $n_k > \bar{n}_k$
($n_k < \bar{n}_k$). It can be shown that this defines a
\textit{total order} among the basis vectors. In particular, it is
clear that $|N,0,\ldots,0 \rangle$ is superior to all other basis
vectors while $|0,0,\ldots,N \rangle$ is inferior to all other basis
vectors.

Having furnished the set of basis vectors with an order structure,
we can now generate all the basis vectors one by one by descending
from the highest one $|N,0,\ldots,0\rangle$. Given a basis vector
$|n_1,n_2,\ldots,n_M \rangle$ with $n_M< N$, we proceed to the next
basis vector inferior to the current one according to the following
rule \cite{k0}:
\begin{spacing}{1.5}
\end{spacing}
Suppose $n_k\neq 0$ while $n_i=0$ for all $k+1 \leq i \leq M-1$,
then the next basis vector is $|\bar{n}_1,\bar{n}_2,\ldots,\bar{n}_M
\rangle$ with

$\bullet$ $\bar{n}_i=n_i$ for $1\leq i \leq k-1$;

$\bullet$ $\bar{n}_{k}=n_k -1$;

$\bullet$ $\bar{n}_{k+1}=N-\sum_{i=1}^k \bar{n}_{i} $ and
$\bar{n}_i=0$ for $i \geq k+2$.
\begin{spacing}{1.5}
\end{spacing}
This procedure will end with the lowest basis vector
$|0,0,\ldots,N\rangle$. Obviously, this algorithm yields a code
involving only a single loop and with all the difficulties
associated with the naive one avoided. Numerically, we store the
basis vectors in a $D\times M$ array \verb"A", with the $v$-th
generated basis vector filled in the $v$-th row of the array. We
will refer to the basis vector in $v$-th row as $| v \rangle$, so
$|v\rangle \equiv |A_{v1},A_{v2},\ldots,A_{vM}\rangle$.

As an example, we enumerate in table~\ref{tab:tc} all the basis
vectors generated with the foregoing algorithm in the case of
$N=M=3$. As for the efficiency of the algorithm, we mention that in
the case of $N=M=13$, it takes about 38 seconds to generate the
$D=5\,200\,300$ basis vectors with our MATLAB code in our desktop
computer \cite{computer}.

\begin{table}[tbh]
\caption{\label{tab:tc} Configurations of the basis vectors
$|n_1,n_2,n_3 \rangle$ with atom number $N=3$ and site number $M=3$.
They are generated recursively according to the algorithm described
in section~\ref{basis gen}.}
\begin{indented}
\item[]\begin{tabular*}{0.3\textwidth}{@{\extracolsep{\fill}} r|c c c}
\hline\hline
$v$ & $n_1$ & $n_2$ & $n_3$ \\
\hline\hline
1 & 3 & 0 & 0 \\
2 & 2 & 1 & 0 \\
3 & 2 & 0 & 1 \\
4 & 1 & 2 & 0 \\
5 & 1 & 1 & 1 \\
6 & 1 & 0 & 2 \\
7 & 0 & 3 & 0 \\
8 & 0 & 2 & 1 \\
9 & 0 & 1 & 2 \\
10 & 0 & 0 & 3 \\
\hline\hline
\end{tabular*}
\end{indented}
\end{table}

\section{Setting up the Hamiltonian matrix}
With all the basis vectors prepared, we are now in the position to
set up the Hamiltonian matrix with respect to this basis. That is,
we are to determine the $D \times D$ matrix \verb"H" corresponding
to the Bose-Hubbard Hamiltonian $\hat{H}$ with
\begin{equation}\label{h}
    \verb"H"_{uv}\equiv\langle u | \hat{H} | v \rangle.
\end{equation}
Here by determining the matrix \verb"H", we do not mean to save it
in the full matrix form in the computer (that will cost memory on
the order of $D^2$), but to figure out all its non-zero elements and
their positions, i.e., their row and column numbers. Actually, as we
will see below, the matrix \verb"H" is extremely sparse with at most
$2M+1$ non-zero elements per column. Therefore, it is appropriate to
store \verb"H" in a certain sparse form, which will require memory
only on the order of $D$. In MATLAB, a sparse matrix is stored in
the coordinate format.

To proceed, we treat the interaction part $\hat{H}_{int}$ and
kinetic part $\hat{H}_{kin}$ of the Hamiltonian separately. The
corresponding matrices are denoted as \verb"H_int" and \verb"H_kin",
respectively. We note that \verb"H_int" and \verb"H_kin" are the
diagonal and off-diagonal parts of \verb"H" respectively. We also
note that this separation is necessary when we want to change the
ratio $U/J$ to study the SF-MI transition. The matrix \verb"H_int"
can be easily done, therefore, we will concentrate on \verb"H_kin".

A general and straightforward but naive method to set up
\verb"H_kin" is to let $u$ and $v$ run over all the integers from 1
to $D$, respectively, and examine the corresponding matrix elements
one by one. This procedure entails computation scale proportional to
$D^2$, and is very inefficient since most checks yield null results.
A clever way out is to ask the question, in each column, which
elements are non-zero? Physically, it is equivalent to ask, given an
arbitrary basis vector $|v\rangle$, if we act $\hat{H}_{kin}$ on it,
which (generally not merely one) basis vectors will appear? To
answer this question, we note that there are $2M$ hopping terms in
$\hat{H}_{kin}$, all in the form of $a_i^\dagger a_j$. These hopping
terms, when acting on a given basis vector, either annihilate it or
change it into another basis vector with some amplitude. In the
latter case, the occupation numbers of the newly generated basis
vector are readily obtained from those of $|v\rangle$. However, this
information is not what we really need. The problem that really
matters is, which basis vector is it among the basis vectors
tabulated in the array \verb"A"? Or more precisely, which row does
it belong to in \verb"A"?

Here we will invoke the so-called hashing technique to fulfill this
aim \cite{knuth,gagliano,our}. The basic idea is to define a tag for
each basis vector, that is, to condense the information of the
vector into a single entity. Thereafter, to see whether two vectors
are the same, rather than comparing their elements one by one, we
only need to see whether their tags are the same. Concretely, the
tag of the $v$-th basis vector is defined by a function $T$,
\begin{equation}\label{j}
    T(v)\equiv T(A_{v1},A_{v2},\ldots,A_{vM}).
\end{equation}
Numerically, this function should be readily evaluated. Moreover,
since we want to identify the basis vectors with their tags, it is
mandatory that different basis vectors have different tags. In other
words, there should be a one-to-one mapping between the rows of
\verb"A" and the elements of the array \verb"T".

A fortunate case is that the tag \verb"T(v)" coincides with $v$. In
this case, by calculating the tag of a basis vector, we know its
rank among all the basis vectors. However, generally it is hard to
find such a function. The compromise is to give up this hope and
impose only the condition that all the tags are different, which is
relatively easy to meet. A candidate of the tag function is
\begin{equation}\label{tag2}
    T(v)=\sum_{i=1}^M \sqrt{p_i} A_{vi},
\end{equation}
with $p_i$ being the $i$-th prime number. This function is linear in
the occupation numbers and are readily calculated. More importantly,
since the $\sqrt{p_i}$'s are radicals of distinct square-free
numbers, they are linearly independent over the rationals
\cite{independent}, and therefore different vectors have different
tags necessarily. An alternative tag function is $T(v)=\sum_{i=1}^M
(\ln p_i) A_{vi}$. By some simple number theory \cite{hardy}, it is
ready to see that this tag function is also a viable one. The tag
function the authors use is of the form (\ref{tag2}) but with
$p_i=100*i +3$, which is easier to program.

Given a basis vector $|v \rangle$ specified by a set of occupation
numbers but with $v$ unknown, we calculate its tag according to
equation~(\ref{tag2}), then search the tag among the array \verb"T"
to locate its position, i.e. the value of $v$. Here another trick is
possible. Originally, the array \verb"T" is unsorted, i.e., the tags
are not arranged in ascending or descending orders according to
their values. To search a given element, the only way is to check
the elements one by one and that will take on average $D/2$ trials
to find out that element. That is a huge work since $D$ can be on
the order of $10^6$. A simple trick saves the workload
significantly. Rather than searching inside an unsorted array, we
had better search inside a sorted array so that we can make use of
the Newton binary method \cite{knuth}. That will take at most
$\log_2D$ trials to find out the target. For $D=2^{20}\simeq
1.05\times 10^6$, it takes at most $20$ trials to locate the target
\cite{though}.

Thus we first sort \verb"T" in ascending (or descending) order with
the quicksort algorithm \cite{knuth,Hoare}. For clarity, we denote
the sorted array as \verb"TSorted". In doing so, we can also prepare
another $D$-element array \verb"ind" which stores the positions of
elements of \verb"TSorted" in the original array \verb"T". More
precisely, \verb"T(ind(i))=TSorted(i)". In MATLAB, this can be done
simply with the code
\begin{verbatim}
    [T,ind]=sort(T)
\end{verbatim}
Here we overwrite the original array \verb"T" with \verb"TSorted".
For those programming with Fortran, the ORDERPACK package by Olagnon
can be used to fulfill the same aim \cite{orderpack}.

We summarize the procedure to establish the matrix \verb"H_kin" as
follows. The non-zero elements are determined column by column.
Given an arbitrary basis vector $|v \rangle$, we apply the hopping
terms $a_i^\dagger a_j$ onto it. If $A_{vj} \geq 1$, we have
\begin{equation}\label{hop basis}
a_i^\dagger a_j |v \rangle= \sqrt{(A_{vi}+1)A_{vj}}| \ldots,
A_{vi}+1,\ldots,A_{vj}-1,\ldots\rangle. \nonumber
\end{equation}
We then calculate the tag $T_r$ of the vector on the right hand side
and search it among the sorted array \verb"T". Suppose
$\verb"T(w)"=T_r$, we then know the resulting basis vector is the
$u=\verb"ind(w)"$ -th one. We have thus found a non-zero element
with coordinates $(u,v)$ and value $-J \sqrt{(A_{vi}+1)A_{vj}}$.
This process is repeated as $v$ runs from $1$ to $D$. Obviously, it
can be parallelized. It is also clear that the overall time cost in
this step is on the order of $M D\log_2D$.

\begin{figure}[tb]
\centering
\includegraphics[bb=155 259 449 580, width=0.35\textwidth]{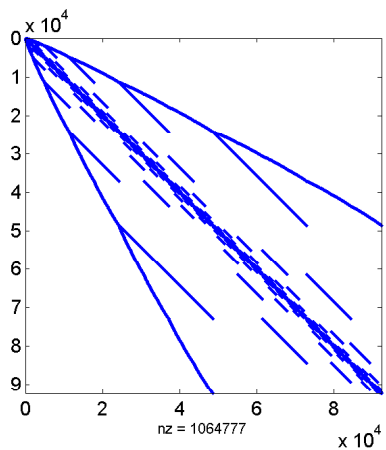}
\caption{\label{fig1} Sparsity pattern of the Hamiltonian matrix
\texttt{H} in the case $N=M=10$. Every spot corresponds to a
non-zero element. The dimension of the Hilbert space is $D=92\,378$
and the number of non-zero elements is $\textrm{nz}=1\,064\,777$. On
average, there are $11.5$ non-zero elements per column.}
\end{figure}

In figure~\ref{fig1}, we plot the sparsity pattern of the
Hamiltonian \verb"H" in the case $N=M=10$. We see that on average
there are only $11.5$ non-zero elements per column, which is four
orders smaller than the dimension $D=92\,378$. We would like to
mention that with our MATLAB code, it takes about 15 seconds to set
up the Hamiltonian matrix and plot the pattern using the code
\verb"spy(H)".

\section{Numerical results}
After preparing the Hamiltonian in a sparse matrix form, we can use
the Lanczos algorithm to compute the ground state and low lying
excited states and their energies. There are some well developed
packages for this purpose and our philosophy is not to reinvent the
wheel. For those programming with Fortran, the ARPACK package
\cite{arpack} by Lehoucq \textit{et al}. is a very useful aid. For
those programming with MATLAB, it is enough to invoke the ``eigs''
command. For instance, the code
\begin{verbatim}
    [Evec,Eval]=eigs(H,2,'sa')
\end{verbatim}
returns the two smallest eigenvalues (the ground state energy and
the first excited state energy) of \verb"H" in the $2\times 2$
diagonal matrix \verb"Eval" and their corresponding eigenvectors
(the ground state and the first excited state) in the $D \times 2$
matrix \verb"Evec". Here we would point out that when executing
``eigs'', MATLAB invokes the very ARPACK package to do the job
\cite{eigs}.

With the ground state on hand, we can then calculate various
quantities to gain some physical insights of the model. One quantity
that is of primary interest is the single-particle density matrix
(SPDM) associated with the many-particle ground state. In the
Wannier state basis, it is defined as
\begin{equation}\label{spdm1}
    \rho^{(1)}_{ij}=\langle G | a_i^\dagger a_j | G \rangle,
\end{equation}
with $1 \leq i,j \leq M$. All one-particle variables, e.g., the
momentum distribution, are captured in the SPDM.

In general, the SPDM is hermitian, semi-positive-definite, and of
trace equal to the particle number. In the present case, the SPDM is
subjected to more constraints. Due to the translation and reflection
invariance of the ground state \cite{ground state}, we have
$\rho^{(1)}_{ij}=\rho^{(1)}_{i+k,j+k}$ for an arbitrary $k$ and also
$\rho^{(1)}_{ij}=\rho^{(1)}_{ji}$. Therefore, the SPDM is real,
symmetric, and cyclic. These good properties reduce the number of
matrix elements to be computed from $M(M+1)/2$ to $[M/2]+1$, where
$[\cdot]$ is the floor function.

\subsection{Condensate fraction}
According to the Penrose-Onsager criterion \cite{penrose}, a
condensate is present if and only if the largest eigenvalue
$\lambda_1$ of $\rho^{(1)}$ is macroscopic, i.e., $f_c=\lambda_1/N$
is on the order of unity and the ratio $f_c$ is called the
condensate fraction. In the non-interacting case, all particles
reside in the lowest Bloch state (a zero momentum state), and the
system is in a pure condensate state with $f_c=1$. As the
interaction is turned on, more and more particles will be kicked
into higher Bloch states and the condensate is said to be depleted.
In the thermodynamic limit ($M$ goes to infinity with $N/M=1$
fixed), there is a critical value \cite{batrouni} ($\simeq 4.65$) of
$U/J$ beyond which $f_c$ vanishes.

\begin{figure}[tb]
\centering
\includegraphics[bb=19 19 315 235,width=0.45\textwidth]{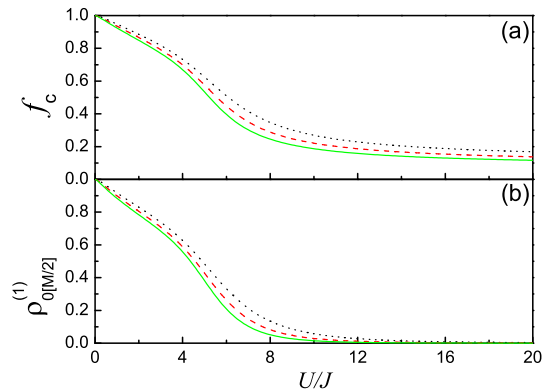}
\caption{\label{fig2}(Colour online) (a) Condensate fraction $f_c$
and (b) correlation $\rho^{(1)}_{0[M/2]}$ as functions of the ratio
$U/J$. In each panel, from up to down, the size of the system is
$N=M=9$, $11$, and $13$, respectively.}
\end{figure}

To gain a picture of this phase transition, we have numerically
calculated the condensate fraction as a function of the ratio $U/J$,
with three different lattice sizes. The results are shown in
figure~\ref{fig2}(a). We see that as $U/J$ increases, the condensate
fraction decreases monotonically. However, the finite size effect is
significant. The condensate fraction is far from being vanishing in
the deep Mott insulator regime ($U/J \gg 4.65$). Actually, since
$\lambda_1 \geq \left(\sum_k \lambda_k\right)/M=N/M$, $f_c$ has a
lower bound $1/M$.

\subsection{Off-diagonal long range order}
The presence of a condensate is also associated with an off-diagonal
long range order \cite{yang}. That is, a condensate is present if
the off-diagonal element of the single particle density matrix
$\rho^{(1)}_{ij}$ converges to a finite value as $|i-j|\rightarrow
\infty$. This is consistent with the Penrose-Onsager criterion.
Actually, converting into the Bloch state representation, the
density matrix takes the form
\begin{eqnarray}
  \tilde{\rho}^{(1)}_{q_1q_2} &=& \langle G | b_{q_1}^\dagger b_{q_2} | G \rangle  \nonumber\\
   &=& \frac{1}{M}\sum_{j_1,j_2}\langle G | a_{j_1}^\dagger a_{j_2} | G\rangle e^{i2\pi(q_1 j_1-q_2 j_2)/M}  \nonumber\\
   &=& \sum\nolimits_{j} \rho^{(1)}_{0j} e^{-i 2\pi q_1 j/M } \delta_{q_1
   q_2}.
\end{eqnarray}
Here in the third line, we have used the cyclicity of $\rho^{(1)}$.
Thus the SPDM is diagonal in the Bloch state representation. Its
eigenvalues coincide with its diagonal elements, and its eigenstates
(called natural orbits) coincide with the Bloch states. It can be
proven that all the elements of $\rho^{(1)}$ are non-negative
\cite{ground state}. Therefore, the largest eigenvalue of the SPDM
is just $\tilde{\rho}^{(1)}_{q_1q_1}$ with $q_1=0$, and is of the
explicit expression
\begin{equation}\label{first}
    \lambda_1=\sum\nolimits_j \rho^{(1)}_{0j}.
\end{equation}
We then see immediately that, if $\rho^{(1)}_{0j}$ decreases
monotonically with $j$, $f_c$ and $\rho^{(1)}_{0j}$ converges to the
same value in the thermodynamical limit.

Thus the phase transition can also be investigated by examining the
behavior of the off-diagonal elements. In figure~\ref{fig2}(b), we
show how the element $\rho^{(1)}_{0[M/2]}$ behaves as $U/J$ varies.
We choose this element because it corresponds to the correlation
between two sites with the largest distance on a circle. Moreover,
as $M$ tends to infinity, the distance $[M/2]$ also tends to
infinity. Comparing with figure~\ref{fig2}(a), we see that the
correlation $\rho^{(1)}_{0[M/2]}$ decreases much faster than $f_c$,
especially, in the deep Mott insulator regime, it does drop to zero.

\subsection{Occupation variance}
\begin{figure}[bth]
\centering
\includegraphics[bb=19 19 287 215,width=0.40\textwidth]{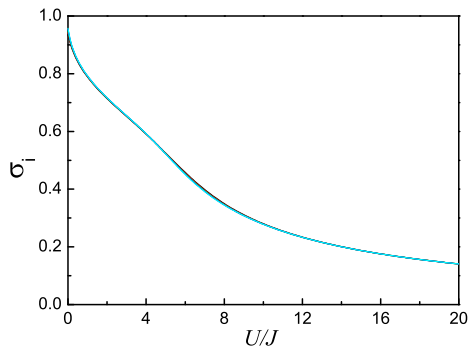}
\caption{\label{fig3} Variance of the occupation number $\sigma_i$
at an arbitrary site $i$ as a function of the ratio $U/J$
\cite{roth}. Actually five different lattice sizes, i.e.,
$N=M=8,9,10,11,12$, are investigated. However, the curves all
collapse onto the same one.}
\end{figure}

In the previous subsections, we see that both the condensate
fraction and off-diagonal elements suffer from a strong finite size
effect. Here, we show that the fluctuation of the occupation number
on one site
\begin{equation}\label{variance}
    \sigma_i=\sqrt{\langle G |\hat{n}_i^2 | G \rangle-\langle G |\hat{n}_i | G \rangle^2}
\end{equation}
is not very sensitive to the size of the lattice \cite{roth}, as
long as $M>>1$. In figure~\ref{fig3}, we show $\sigma_i$ as a
function of $U/J$ for five different lattice sizes. The difference
between the curves are hardly visible. This indicates that the curve
has already converged to its value in the infinite lattice limit.

This suggests that the SF-MI transition can not be tracked in any
local variables. This is why we do not fix the distance between the
two sites when calculating the off-diagonal element in the
proceeding subsection. By the Hellmann-Feynman theorem, we have
$\partial E_G /\partial U= \langle \partial H/\partial U
\rangle=M\sigma_i^2/2$. Our numerical result for $\sigma_i$ suggests
that the ground state energy is a smooth function of $U$.

\section{Conclusion and discussion}
Exact \diag\ is simple conceptually but never trivial in
programming. Taking the \bh\ model as a working example, we have
illustrated the architecture of numerical exact \diag. Some
essential tricks, namely, ordering, enumerating, hashing, sorting,
and searching, were explained in detail. These tricks are believed
to be quite general and can be adopted in many other situations
\cite{bertsch}.

For example, we show how the idea of ordering can save computation
if we want to work within the subspace $\mathcal {H}_0$. To impose
the condition $K=0$, we had better work with $\hat{H}$ in
equation~(\ref{h momentum}) \cite{k0}. This time, it is the
interaction part
\begin{equation}\label{hint}
    \hat{H}_{int}=\frac{U}{2M}\sum_{q_1,q_2}\sum_{q_3,q_4} b_{q_1}^\dagger b_{q_2}^\dagger b_{q_3}
    b_{q_4}
\end{equation}
that costs most effort. Here the condition $q_1+q_2 \equiv q_3+q_4
\pmod M$ is taken implicitly. There are exactly $M^3$ terms in the
sum. By noting
$[b_{q_1},b_{q_2}]=[b^\dagger_{q_3},b^\dagger_{q_4}]=0$, we have
\begin{equation}\label{hint2}
 \hat{H}_{int}=\frac{U}{2M}\sum_{q_1 \geq q_2}\sum_{q_3\geq q_4} B_{q_1 q_2}^\dagger  B_{q_3
 q_4},
\end{equation}
where $B_{q_1 q_2}^\dagger=(2-\delta_{q_1 q_2}) b_{q_1}^\dagger
b_{q_2}^\dagger$ is defined for $q_1 \geq q_2$. We can reduce the
computation further by making use of the hermicity of
$\hat{H}_{int}$. We define $(q_1 q_2)=Mq_1+ q_2$. The terms
$B^\dagger_{q_1 q_2}$ are ordered according to their tags $(q_1
q_2)$. We then rewrite (\ref{hint2}) as
\begin{equation}
\fl  \hat{H}_{int}=\frac{U}{2M}\bigg[ \sum_{{(q_1 q_2)=(q_3 q_4)}}
B_{q_1 q_2}^\dagger  B_{q_3
 q_4}
  +\sum_{{(q_1 q_2)>(q_3q_4)}} B_{q_1 q_2}^\dagger  B_{q_3
 q_4} +\sum_{{(q_1 q_2)<(q_3q_4)}} B_{q_1
q_2}^\dagger  B_{q_3
 q_4}\bigg].
\end{equation}
The third term is hermitian conjugate to the second
one. Thus in setting up the matrix corresponding to $\hat{H}_{int}$,
we only need to consider the non-zero elements due to the second
term, those due to the third term are then determined automatically.
Overall, in the case of $N=M=10$, the number of terms need to be
considered is reduced from 1000 to 180.

The readers are encouraged to convert the procedure described in
this paper into codes and explore the interesting physics in the
\bh\ model, which is surely far from being exhausted in the present
paper. For example, we have shown how to study the condensate
fraction as a function of the ratio $U/J$. On this basis, an
immediately accessible problem is then how the superfluidity density
varies with $U/J$. The subtle relation between condensation and
superfluidity can then be investigated. For more details, see
\cite{roth}. We would like to mention that the coding does not cost
much effort. It takes no more than 100 lines in MATLAB, and is very
efficient. For the $N=M=12$ case, it takes around 4 minutes to set
up the Hamiltonian matrix and 1 minute to solve the ground state on
our desktop computer. Note that the dimension is $D=1\,352\,078$. By
working in the subspace $\mathcal {H}_0$, we have successfully
performed exact \diag\ for a system as large as $N=M=13$ on our
computer. Systems with larger sizes may be investigated by working
in the $\mathcal {H}_0^+$ subspace. Therefore, the \bh\ model may
serve as a good topic for teaching exact \diag\ in an undergraduate
or graduate course of computational physics.

\ack

The authors are grateful to G.~F. Bertsch and T. Papenbrock for
their helpful advices and to E.~H. Lieb for his valuable comments.


\begin{thebibliography}{59}
\bibitem{identify}
Due to the periodic boundary condition, the site index $0$ is
identified with $M$.
\bibitem{ground state}
For $J>0$, due to the connectedness of the lattice and the
Perron--Frobenius theorem, the ground state $|G\rangle$ must be
non-degenerate and all the expansion coefficents in terms of the
Fock states (\ref{ocuupation}) are real and of the same sign.
Therefore, we have $\langle G |e^{i\hat{K}}| G \rangle > 0$. Since
$| G\rangle$ is an eigenstate of the unitary operator
$e^{i\hat{K}}$, we have $e^{i\hat{K}}| G \rangle  = | G \rangle $ .
That is, the ground state is translationally invariant. Moreover, it
is easy to show that it is also invariant under reflection.
Therefore, $|G\rangle$ constitutes the identity representation of
the group $D_M$.
\bibitem{k0}
The same rule can be used to generate the basis vectors in the
subspace $\mathcal {H}_0$. The space $\mathcal {H}$ is spanned by
$\{ |m_0,m_1,\ldots,m_{M-1}\rangle\}$ which are defined as
$b_q^\dagger b_q |m_0,m_1,\ldots,m_{M-1}\rangle= m_q
|m_0,m_1,\ldots,m_{M-1}\rangle$ with $\sum_q m_q=N$. These states
are generated one by one but only those with $\sum_q q m_q \equiv 0
\pmod {M}$ will be taken into $\mathcal {H}_0$. The subspace
$\mathcal {H}_0^+$ of $\mathcal {H}_0$ is spanned by
$\{\frac{1}{\sqrt{2}}(|m_0,m_1,\ldots,m_{M-1}\rangle+|m_0,m_{M-1},\ldots,m_{1}\rangle)
\}$.
\bibitem{our}
We should mention that we use the hashing trick not in the standard
way, as in \cite{knuth,gagliano}. In our approach, we are free of
the troublesome collision problem in the usual way of hashing. The
coding is thus made much easier and more suitable for pedagogical
use.
\bibitem{though}
Though on average it takes 19 trials.
\end{thebibliography}
\end{document}